\documentstyle[preprint,aps,epsf,floats]{revtex}    
\begin{document}
\everymath={\displaystyle}
\def\bra{\langle}
\def\ket{\rangle}
\def\ketc{\rangle^\ast}
\def\bq{\begin{equation}}
\def\eq{\end{equation}}
\def\ba{\begin{eqnarray}}
\def\ea{\end{eqnarray}}
\def\O{{\cal{O}}}
\def\as{\alpha_s}
\def\smin{s_{\mbox{\scriptsize min}}}
\def\lsimfig{\mathrel{\raise.2ex\hbox{$<$}\hskip-.8em\lower.9ex\hbox{$\sim$}}}
\def\gsimfig{\mathrel{\raise.2ex\hbox{$>$}\hskip-.8em\lower.9ex\hbox{$\sim$}}}
\def\lsim{\roughly<}
\def\gsim{\roughly>}
\def\mboxsc#1{\mbox{\scriptsize #1}}
\tighten
%
%
\preprint{
\font\fortssbx=cmssbx10 scaled \magstep2
\hbox to \hsize{
\hfill$\vtop{
                \hbox{\bf TTP97-30}
                \hbox{\bf hep-ph/9708231}
                \hbox{August 1997}
                \hbox{}
                \hbox{}}$}
}
\title{Effects of Jet Azimuthal Angular Distributions \protect\\on 
       Dijet Production Cross Sections in DIS}
\author{Erwin Mirkes and  Stefan Willfahrt\\[3mm]}
\address{Institut f\"ur Theoretische Teilchenphysik, 
         Universit\"at Karlsruhe,\\ D-76128 Karlsruhe, Germany}
\maketitle
\vspace*{2cm}
\begin{abstract}
A Monte Carlo study of the azimuthal angular distribution
around the virtual boson-proton beam axis for dijet events 
in DIS at HERA is presented.
In the presence of typical acceptance cuts on the jets in the
laboratory frame, the azimuthal distribution
is dominated by  kinematic effects rather than the
typical $\cos\phi$ and $\cos 2\phi$ dependence 
predicted by the  QCD matrix elements.
This implies that the $\phi$ dependent part 
of the QCD matrix elements contributes even to the dijet  production 
cross section. Neglecting this $\phi$ dependence 
leads to an error of about 5-8\% in the production cross
section for typical acceptance cuts in the laboratory frame.
We also present first NLO results on the $\phi$-decorrelation
of the jets through NLO effects.
\end{abstract}
\thispagestyle{empty}
%
%
%
%
%
\newpage
Jet production in deep inelastic scattering
(DIS) is an important laboratory for testing
our understanding of perturbative QCD.
Good event statistics allow for precise measurements
and thus for a variety of tests of our understanding of QCD dynamics. 
For quantitative studies one clearly needs to compare data
with calculations which include next-to-leading order (NLO) QCD 
corrections.
The importance of keeping the full helicity structure 
in such a calculation is demonstrated in this letter
for the example of dijet production in DIS.

Dijet production in neutral current (NC)  DIS
lepton proton scattering 
proceeds via the exchange of an
intermediate vector boson $\gamma^\ast/Z$.
We denote the $\gamma^\ast/Z$  momentum by $q$,
its absolute square by $Q^2$, and use the standard scaling variables 
$x={Q^2}/({2P\cdot q})$ and $y={P\cdot q}/{P\cdot l}$.
In Born approximation, the subprocesses  
\begin{eqnarray}
e(l)+q(p_0)&\rightarrow& e(l^\prime) + q(p_1) + g(p_2)
\label{qtoqg}\\
e(l)+g(p_0)&\rightarrow& e(l^\prime) + q(p_1) + \bar{q}(p_2)
\label{gtoqqbar}
\end{eqnarray}
contribute to the NC  two-jet cross section.
NLO effects will be discussed later.
{\it In the absence of jet cuts in the laboratory frame},
the full leading order (LO) QCD matrix elements predict a typical
azimuthal angular distribution of the jets of the form \cite{old,herai}
\begin{equation}
\frac{d\sigma}{d\cos\phi}=A+B\cos\phi+C\cos 2\phi
\label{dphi}
\end{equation}
where $\phi$ denotes the azimuthal angle of the jets around the
virtual boson direction. This angular distribution is determined
by the gauge boson polarization:
the coefficients $A,B,C$ in Eq.~(\ref{dphi}) are linearly
related to the nine polarization density elements $h_{mm^\prime}$ 
of the exchanged gauge boson:
\begin{equation}
h_{mm^{\prime}}= \epsilon_{\mu}^{\ast}(m)H^{\mu\nu}\epsilon_{\nu}(m^{\prime})
\hspace{1cm}
(m,m^{\prime}=+,0,-)
\end{equation}
where
\begin{equation}
\epsilon_{\mu}(\pm) = \frac{1}{\sqrt{2}}(0;\pm1,-i,0)
\hspace{1cm}
\epsilon_{\mu}(0) = (0;0,0,1)
\end{equation}
are the polarization vectors  of the exchanged boson in the
hadronic ({\it i.e.} boson-proton)
center of mass frame (HCM)
and $H^{\mu\nu}$ denotes the hadronic tensor for the
contributing QCD processes in Eqs.~(\ref{qtoqg},\ref{gtoqqbar}).
The coefficient $A$ in Eq.~(\ref{dphi}) can be expressed in terms of the
diagonal density matrix elements $h_{00}, h_{++}, h_{--}$,
the coefficient $B$ in terms of transverse-longitudinal interference
matrix elements $h_{+0},h_{0+},h_{-0},h_{0-}$
and the coefficient $C$ in terms of transverse interference
density matrix elements $h_{+-},h_{-+}$
\cite{tom}.
Technically, the coefficient $A$ can be calculated from the
hadronic tensor in the one photon exchange case
by the two covariant projections
$g_{\mu\nu}$ and $p_{0\,\mu}p_{0\,\nu}$ on the
hadronic tensor, whereas projections
defined with final state hadronic momenta are necessary 
to calculate $B$ and $C$ \cite{herai}.

The NLO $eP\rightarrow n$ jets event generator MEPJET \cite{letter}
is used for the subsequent numerical studies in this paper.
The corresponding cross sections and distributions 
are calculated by  evaluating the corresponding one-loop
and tree-level helicity amplitudes and therefore,
the MEPJET program allows for the calculation
of all possible jet-jet and jet-lepton
correlations in NLO, including the $\phi$ dependence 
in Eq.~(\ref{dphi}).

The following numerical studies for $e^+P$ scattering
with $\gamma/Z$ exchange are based on
MRSR1 \cite{mrsr1} parton distribution functions 
with the two loop formula for the strong coupling constant
and $n_f=5$.
The renormalization and the factorization scales  are set to
$\mu_R=\mu_F= \langle k_T^B(j)\rangle$, the average $k_T^B$
of the jets, which appears to be
the ``natural'' scale for jet production in DIS \cite{rheinsberg}.
Here, $(k_T^{B}(j))^2$ is defined by $2\,E_j^2(1-\cos\theta_{jP})$
in the Breit frame, where the subscripts $j$ and $P$ 
denote the jet and proton, respectively.
The lepton and hadron beam energies are 27.5 and 820 GeV, respectively
and events are selected in the following kinematical region:
40~GeV$^2<Q^2<2500$ GeV$^2$, $0.04 < y < 1$, 
and $E(e^\prime)>10$~GeV, where $E(e^\prime)$ is the
energy of the scattered lepton.
Unless stated otherwise, no
cuts on the pseudo-rapidity $\eta^{\mboxsc{lab}}=-\ln\tan(\theta/2)$
or the transverse momentum on the jets in the lab frame
are imposed.

\begin{figure}[tb]
\vspace*{0.0in}            
\begin{picture}(0,0)(0,0)
\includegraphics{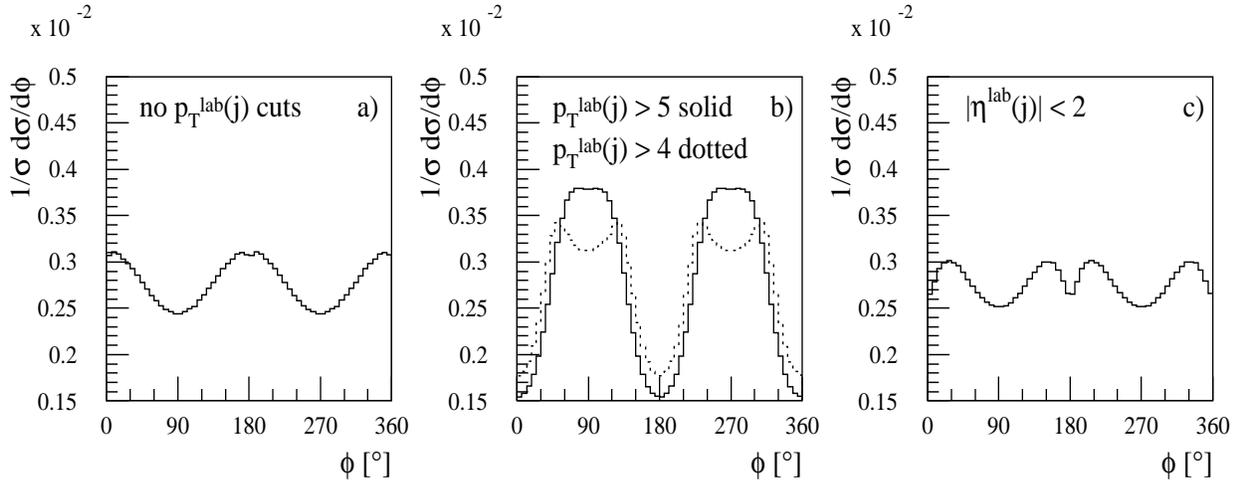}
\end{picture}
\vspace{6.5cm}
\caption{
a) Normalized $\phi$ distribution in LO  around the virtual boson
direction of jets in the cone scheme defined in
the HCM with $p_{T}^{\protect\mboxsc{HCM}}(j)>5$ GeV.
The $\phi$ integrated dijet cross section is 1465pb;
b) 
same as a) but but with an additional cut of $p_T^{\protect\mboxsc{lab}}(j)>5$ 
(solid)
($p_T^{\mboxsc{lab}}(j)>4$ (dotted)) GeV on the jets.
The $\phi$ integrated dijet cross section is 945 pb
(1146 pb);
c) 
same as a) but with an additional 
$|\eta^{\protect\mboxsc{lab}}(j)|<2$ cut on the jets.
The $\phi$ integrated dijet cross section is 1087 pb.
All results are given in LO with $\gamma^\ast$ and $Z$ exchange.
}
\label{fig1}
\end{figure}

Without an (experimental) separation
of a quark, anti-quark or gluon jet, the $\cos\phi$ term
in Eq.~(\ref{dphi}) is not observable and thus 
has to be averaged out.
The resulting LO $\cos 2\phi$ dependence of dijet events
is shown in Fig.~\ref{fig1}a for a cone scheme 
defined in the HCM with radius $R=1$
and $p_T^{\mboxsc{HCM}}(j)>5$~GeV.
We find  very similar results for jets 
in the $k_T$ algorithm  \cite{kt} implemented in the Breit frame
with $E_T^2=40$  GeV$^2$ and $y_{\mboxsc{cut}}=1$.
Note that the HCM and the Breit frame are related
by a boost along the boson-proton direction which does not change
the $\phi$ dependence in Eq.~(\ref{dphi}).
One observes a sizable $\cos 2\phi$ dependence
(after averaging over $\cos\phi$) in Fig.~\ref{fig1}a
as  expected from Eq.~(\ref{dphi}).
The size of the $\phi$ dependence in this normalized distribution
is rather insensitive on the 
$p_{T\,\mboxsc{min}}^{\mboxsc{HCM}}$ 
(or $E_T$ in the $k_T$ scheme)  requirements on the jets.
Averaging (integrating) over $\phi$ implies that
the  coefficients $B$ and $C$ in Eq.~(\ref{dphi}) 
do not contribute to the dijet production cross section
in this case.

Fig.~\ref{fig1}b shows the $\phi$ distribution of the jets 
defined in the cone scheme and cuts as in Fig.~1a,
but with an additional  $p_T^{\mboxsc{lab}}(j)>5$ (solid)
($p_T^{\mboxsc{lab}}(j)>4$ (dotted)) GeV cut on the jets in the laboratory frame.
Only 65\% (78\%) of the events pass the additional 
$p_T^{\mboxsc{lab}}(j)$ cuts.
These cuts have a dramatic effect on the shape of the
$\phi$ distribution. The shape of the distribution is
now governed by the kinematics of the surviving events.
The $p_T^{\mboxsc{lab}}(j)$ cut introduces a strong $\phi$ dependence.
The ``kinematical'' $\phi$ distribution in Fig.~\ref{fig1}b is very different
from the ``dynamical'' $\cos 2\phi$ distribution in Fig.~\ref{fig1}a.
The only remaining vestiges of the gauge boson polarization effects in the
$\phi$ distribution are the dips at $\phi=90^\circ$ and
$270^\circ$ in the dotted curve in Fig.~\ref{fig1}b.
Imposing cuts on the pseudo-rapidities $|\eta^{\mboxsc{lab}}(j)|$
of the jets in the laboratory frame has also a large
effect on the shape of the $\phi$ distribution 
as shown in Fig.~\ref{fig1}c.

A consequence of this kinematical $\phi$ dependence is, 
that the $\phi$ dependent coefficients $B$ and $C$
in Eq.~(\ref{dphi}) now also contribute
to the dijet {\it production}
cross section after integration over $\phi$.
We have analyzed this effect by comparing the dijet cross
section using the exact matrix elements (including the $\phi$ dependent 
coefficients in Eq.~(\ref{dphi})) with the resulting
dijet cross section where the
coefficients $B$ and $C$ in the matrix elements
in Eq.~(\ref{dphi}) are neglected.
The results are shown in table~\ref{tab1} for various
$p_T^{\protect\mboxsc{lab}}(j)$  cuts on the jets.
We find that  the dijet cross section based on the
approximate matrix elements is larger compared
to the correct result.
The error depends on the 
$p_T^{\protect\mboxsc{lab}}(j)$  cuts and can reach about 6.5 \%.
The error from the approximate matrix elements 
in the presence of pseudo-rapidity cuts on the jets in the lab frame
is shown in Table~\ref{tab2}.
A combination of $p_T^{\protect\mboxsc{lab}}(j)$
and $|\eta^{\mboxsc{lab}}(j)|$ cuts can lead to even larger effects.

\begin{table}[tb]
\caption{
Effects of the jet azimuthal angular distribution on the
dijet production cross section in DIS as a function
of $p_T^{\protect\mboxsc{lab}}(j)$ cuts on the jets.
$\sigma^{\protect\mboxsc{exact}}$ ($\sigma^{\protect\mboxsc{approx}}$) is based
on the exact (approximate) matrix elements including (without) the
coefficients $B$ and $C$ in Eq.~(\protect\ref{dphi}).
Jets are defined in the HCM with radius $R=1$ and
and $p_T^{\protect\mboxsc{HCM}}(j)>5$~GeV.
Additional parameters are explained in the text.
Similar results are found for jets defined in the $k_T$ scheme.
}
\label{tab1}
\vspace{2mm}
\begin{tabular}{llll}
       lab frame cut
     & $\sigma^{\mboxsc{exact}}$[2-jet]
     & $\sigma^{\mboxsc{approx}}$[2-jet]
     & $\Delta\sigma=[\sigma^{\mboxsc{approx}}-\sigma^{\mboxsc{exact}}]/
       \sigma^{\mboxsc{exact}}$  \\
\hline\\[-3mm]
$p_T^{\mboxsc{lab}}(j)>0$
     & 1465    pb
     & 1465    pb  
     & 0  \%   \\
$p_T^{\mboxsc{lab}}(j)>2$
     & 1390    pb
     & 1405    pb  
     & 1.1 \%   \\
$p_T^{\mboxsc{lab}}(j)>4$
     & 1145    pb
     & 1195    pb  
     & 4.4 \%   \\
$p_T^{\mboxsc{lab}}(j)>6$
     & 689    pb
     & 733    pb  
     & 6.4 \%    \\
$p_T^{\mboxsc{lab}}(j)>8$
     & 348    pb
     & 370    pb  
     & 6.3 \%   \\
$p_T^{\mboxsc{lab}}(j)>10$
     & 187    pb
     & 198    pb  
     & 5.9\%   \\
\end{tabular}
\end{table}
\begin{table}[t]
\caption{
Same as table~\protect\ref{tab1} as a function
of $|\eta^{\mboxsc{lab}}(j)|$ cuts on the jets.
}
\label{tab2}
\vspace{2mm}
\begin{tabular}{llll}
       lab frame cut
     & $\sigma^{\mboxsc{exact}}$[2-jet]
     & $\sigma^{\mboxsc{approx}}$[2-jet]
     & $\Delta\sigma=[\sigma^{\mboxsc{approx}}-\sigma^{\mboxsc{exact}}]/
       \sigma^{\mboxsc{exact}}$  \\
\hline\\[-3mm]
$|\eta^{\mboxsc{lab}}(j)|<5$
     &   1465  pb
     &   1465  pb  
     &     0 \%   \\
$|\eta^{\mboxsc{lab}}(j)|<3.5$
     &   1437  pb
     &   1439  pb  
     &     0.15\%  \\
$|\eta^{\mboxsc{lab}}(j)|<2.5$
     &   1282  pb
     &   1288   pb  
     &    0.5\%   \\
$|\eta^{\mboxsc{lab}}(j)|<1.5$
     &   801  pb
     &   820  pb  
     &   2.4\%     \\
$|\eta^{\mboxsc{lab}}(j)|<0.5$
     &   131  pb
     &   141  pb  
     &   7.6\%     \\
\end{tabular}
\end{table}

Let us now discuss some NLO effects on the dijet 
azimuthal angular distribution
in $e^+P$ scattering with $\gamma/Z$ exchange,
where the full helicity structure is kept.
The NLO $\O(\as^2)$ 2-jet cross section receives contributions
from the one-loop corrections to the Born processes
in Eqs.~(\ref{qtoqg}) and (\ref{gtoqqbar}) 
and from the integration over the unresolved region
(defined by a given jet algorithm) of the 3-parton final
state tree level matrix  elements
\begin{eqnarray}
e(l)+q(p_0)&\rightarrow& e(l^\prime) + q(p_1) + g(p_2) + g(p_3)
\label{qtoqgg} \\[1mm]
e(l)+q(p_0)&\rightarrow& e(l^\prime) + q(p_1) + q(p_2) + \bar{q}(p_3)
\label{qtoqqqbar} \\[1mm]
e(l)+g(p_0)&\rightarrow& e(l^\prime) + q(p_1) + \bar{q}(p_2) + g(p_3)
\label{gtoqqbarg}
\end{eqnarray}
and the corresponding antiquark processes with $q\leftrightarrow \bar{q}$.

The NLO corrections introduce a new (numerically small)
additional $\sin\phi$ and $\sin2\phi$ dependence in the NC
cross section in Eq.~(\ref{dphi})
through the imaginary parts of the 
one-loop contributions \cite{hagiwara}.

In addition, the tree level contributions to dijet production in 
Eqs.~(\ref{qtoqgg}-\ref{gtoqqbarg}) imply that the two jets
are no longer necessarily back-to-back in $\phi$ like in LO, {\it i.e.}
one expects $\Delta\phi=|\phi(\mbox{jet 1})-\phi(\mbox{jet 2})|
\neq 180^\circ$.
A deviation from $\Delta\phi=180^\circ$ can arise 
for example in the cone scheme 
if one of the three final state partons 
is well separated in $\Delta R$ from the other two partons
but does not pass the acceptance cuts (like the
$p_T^{\mboxsc{HCM}}(j)>5$ GeV cut),
whereas the remaining two partons passes  all
jet requirements. Thus the event will be accepted
as a two-jet event where the jets are, however, no longer balanced in $\phi$.

\begin{figure}[h]
\vspace*{1in}            
\begin{picture}(0,0)(0,0)
\includegraphics{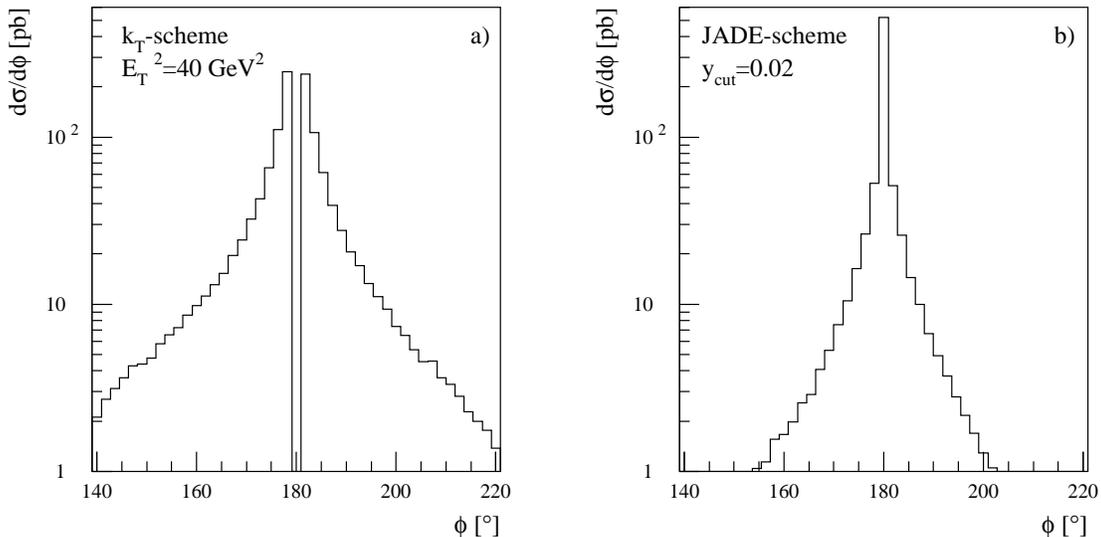}
\end{picture}
\vspace{6cm}
\caption{
a) $\Delta \phi$ distribution in NLO for jets defined in the
$k_T$ scheme with $E_T^2=40$ GeV$^2$. 
Similar results are found for jets in the cone scheme
defined in the HCM.
b) 
same as a) for the JADE scheme and $y_{\protect\mboxsc{cut}}=0.02$.
}
\label{fig2}
\end{figure}
The effect is shown in Fig.~\ref{fig2} for 
the $k_T$ scheme defined in the Breit frame
with $E_T^2=40$ GeV$^2$ (Fig.~\ref{fig2}a) and
the JADE scheme \cite{jade,herai}
defined in the lab frame with $y_{\mboxsc{cut}}=0.02$
(Fig.~\ref{fig2}b).
In the latter case, jets are required to have a minimum transverse momentum
of 2 GeV in the HCM and the lab frame.
The corresponding NLO (LO) cross sections are
1350 pb (1240 pb) in the $k_T$ scheme and 1570 pb (970 pb) in the JADE scheme.

Fig.~\ref{fig2} illustrates that the decorrelation effect through the
NLO corrections depends strongly on the chosen jet algorithm.
The decorrelation is larger in the $k_T$ scheme (or a cone scheme)
than in the JADE scheme.
Note that the central bin around 180$^\circ$ 
in Fig.~\ref{fig2}a has a negative weight. This
shows that the fixed NLO predictions 
are not infrared save for $\Delta\phi$ close to 180$^\circ$.
This effect is caused by the negative contributions from the
virtual corrections, which contribute only to this bin
due to the Born kinematics. 
For the JADE scheme the negative contributions in the central bin
are already overcompensated by the positive tree-level contributions in 
Eqs.~(\ref{qtoqgg}-\ref{gtoqqbarg}).
Since the decorrelation effect is larger in Fig.~\ref{fig2}a,
one has either to choose wider bins in $\Delta\phi$ for the $k_T$ scheme
to arrive at a positive result in the central bin 
or alternatively one would have to use resummation techniques to
obtain a reliable perturbation expansion close to $\Delta\phi$=180$^\circ$.

Finally, the small asymmetry in the $\Delta\phi$ decorrelation
in Fig.~\ref{fig2} is caused
by our fixed ordering of jet~1 and jet~2. 
The distributions  would be perfectly symmetric
without a separation of a quark, anti-quark or gluon jet.

\vspace*{1cm}
\noindent{\bf To summarize:} 
The azimuthal angular distribution of dijet events around the
boson proton direction has been discussed. In the presence of cuts on the
jets in the lab frame, the angular distribution is dominated by kinematic 
effects and the residual dynamical effects from the gauge boson
polarization are small. The kinematic effects imply that the
non-diagonal helicity density matrix elements of the exchanged boson,
which are fully taken into account in the NLO event generator
MEPJET, contribute also to the dijet production cross section.
First studies of the dijet angular decorrelation through NLO corrections
show a fairly strong dependence on the chosen jet algorithm.
%

\end{document}